\documentclass{elsart}
\usepackage{epsfig}
\usepackage{lineno}

\begin{document}
\runauthor{Marco Battaglia}
\begin{frontmatter}
\title{Characterisation of a Thin Fully-Depleted SOI Pixel Sensor with Soft X-ray Radiation}
\author[LBNL,UCSC]{Marco Battaglia,}
\author[INFN]{Dario Bisello,}
\author[LBNL]{Richard Celestre,}
\author[LBNL]{Devis Contarato,}
\author[LBNL]{Peter Denes,}
\author[INFN]{Serena Mattiazzo,}
\author[LBNL]{Craig Tindall}
\address[LBNL]{Lawrence Berkeley National Laboratory, 
Berkeley, CA 94720, USA}
\address[UCSC]{Santa Cruz Institute of Particle Physics, University of California 
at Santa Cruz, CA 95064, USA}
\address[INFN]{Dipartimento di Fisica, Universit\'a di Padova and INFN,
Sezione di Padova, I-35131 Padova, Italy}

\begin{abstract}
This paper presents the results of the characterisation of a back-illuminated pixel sensor 
manufactured in Silicon-On-Insulator technology on a high-resistivity substrate with soft 
X-rays. The sensor is thinned and a thin Phosphor layer contact is implanted on the back-plane. 
The response to X-rays from 2.12 up to 8.6~keV is evaluated with fluorescence radiation at the 
LBNL Advanced Light Source.  
\end{abstract}
\end{frontmatter}

\typeout{SET RUN AUTHOR to \@runauthor}


\section{Introduction}

The availability of the ROHM Lapis Semiconductor (former OKI Industries) Silicon on Insulator (SOI) process 
with an handle wafer of moderate resistivity and contacts through the buried oxide layer has enabled a 
significant R\&D on monolithic Si pixel sensors for charged particle tracking and imaging. 
The SOI technology has a number of potential advantages compared to bulk CMOS processes for the 
fabrication of pixel sensors. Past the first proof of principle of beam particle detection with an 
SOI pixel sensor~\cite{Battaglia:2007eq}, the R\&D had to solve the back-gating effect, which limited 
the practical depletion voltage and thus the depleted thickness.
The use of a buried p-well (BPW) to protect the CMOS electronics from the potential on the handle wafer 
in the sensor substrate has successfully solved this problem and SOI pixels have demonstrated full 
functionality up to 90~V and above~\cite{soi2, soi3, soibeam}, corresponding to a depleted 
thicknesses $\ge$130~$\mu$m. An SOI pixel sensor is potentially well suited for applications in X-ray 
imaging and reference~\cite{soiX} discusses tests of a prototype chip for application in hard X-ray 
imaging spectroscopy on future astronomical satellites.

In this paper we present the results of the characterisation of an SOI pixel sensor with soft X-rays in 
back-illumination. For this application the sensor is thinned to 70~$\mu$m, to ensure full depletion, 
and the back-plane post-processed to create a thin entrance window. Results obtained with the same thin, 
back-processed SOI sensor on an high energy hadron beam to study its performance in particle vertex 
tracking for accelerator particle physics are presented in a companion paper~\cite{soi4}.

\section{Sensor back-plane post-processing}

The pixel sensor prototype has simple 3T analog pixels arrayed on a 13.75~$\mu$m pitch. It has been 
designed at LBNL and produced in the OKI 0.2~$\mu$m SOI process. The handle wafer is of Czochralski (CZ) 
type and has a nominal resistivity of 700~$\Omega$$\cdot$cm and buried oxide thickness of 200~nm.  
This chip has already been characterised both in the lab with laser beams of different wavelengths 
and on a beam-line at the CERN SPS using 200~GeV $\pi^-$. These tests demonstrated that pixel cells 
with BPW extending below either the diode or the transistors are not affected by the back-gating 
effect up to voltages $\ge$70~V. The cluster signal-to-noise ratio for minimum ionising particles was 
measured to be 55, the particle detection efficiency $\ge$0.98 and the single point resolution 
(1.12$\pm$0.03)~$\mu$m, for $V_d \ge$ 50~V~\cite{soibeam}.  

These sensors have a breakdown voltage of $\sim$130~V, which prevents full depletion of their 260~$\mu$m 
full thickness, as provided by the foundry.  Therefore, a set of sensor chips 
has been back-thinned to 70~$\mu$m using a commercial grinding technique~\cite{aptek}, which has been already 
successfully employed for back-thinning CMOS Active Pixel Sensors~\cite{backthin}. After thinning, the sensor 
leakage current increases by more than one order of magnitude, due to the defects generated on the back-plane 
by the grinding process. 

\begin{figure}[h!]
\begin{center}
\includegraphics[width=7.5cm,clip=]{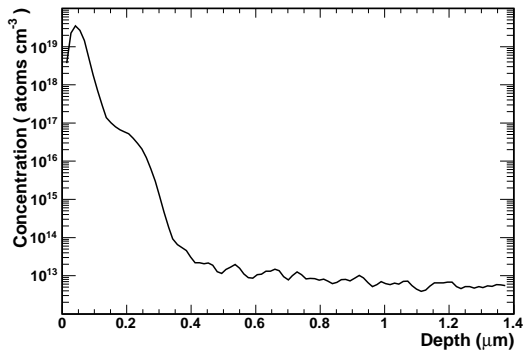}
\end{center}
\caption{P concentration profile on the back-plane from SRA analysis of a post-processed sensor chip.}
\label{fig:sra}
\end{figure}

The thinned sensors are post-processed to create a thin entrance window on the 
back-plane and anneal the crystal damage from the thinning. A thin Phosphor layer contact is implanted at 
33~keV using a cold process at -160$^{\circ}$C. The dose is adjusted to obtain an amorphous layer of Si, 
which favours crystal re-growth. After P implant the chip is annealed at 500$^{\circ}$C for ten minutes 
in Nitrogen atmosphere. The thickness of the P implantation is measured using spreading resistance analysis 
(SRA) on a chip. 
The result is shown in Figure~\ref{fig:sra} and indicates that the P implant extends to a depth of 
$\simeq$0.4~$\mu$m from the back-plane surface, with the highest concentration in the first 0.2~$\mu$m.
This process provides the sensors with a thin entrance window on the back-plane ensuring in principle good 
sensitivity to X-rays down to $\sim$1.5~keV and to electrons down to $\sim$9~keV.
After back-plane post-processing the leakage current is reduced to values comparable to, or lower than, those 
obtained on un-processed chips~\cite{soi4} and the sensors are functional with a total single pixel noise of 
(95$\pm$6)~$e^-$ ENC, which is consistent with that of (83$\pm$8)~$e^-$ measured on thick, un-processed sensors, 
and can be fully depleted. 

\section{Experimental set-up and results}

The data acquisition of the SOI sensor is performed with the setup discussed in~\cite{soibeam,Battaglia:2009daq}. 
Most of the measurements reported here have been performed with the chip clocked at 6.25~MHz, corresponding 
to a read-out time of 160~ns per pixel. Data sparsification is performed on-line in the DAQ PC using a custom 
{\tt Root}-based~\cite{Brun:1997pa} program. The matrix of pedestal subtracted data is scanned for seed pixels 
with signal exceeding a preset threshold in noise units. For each seed, the 7$\times$7 pixel matrix centred around 
the seed position is selected and stored on file. Signal clusters are reconstructed offline by applying a double 
threshold method on the matrix of pixels selected around a candidate cluster seed. Clusters are required to have a 
seed pixel with a signal-to-noise ratio, S/N, of at least 5.0 and the neighbouring pixels with a S/N in excess of 2.5. 

\begin{figure}[h!]
\begin{center}
\begin{tabular}{cc}
\includegraphics[width=6.5cm,clip=]{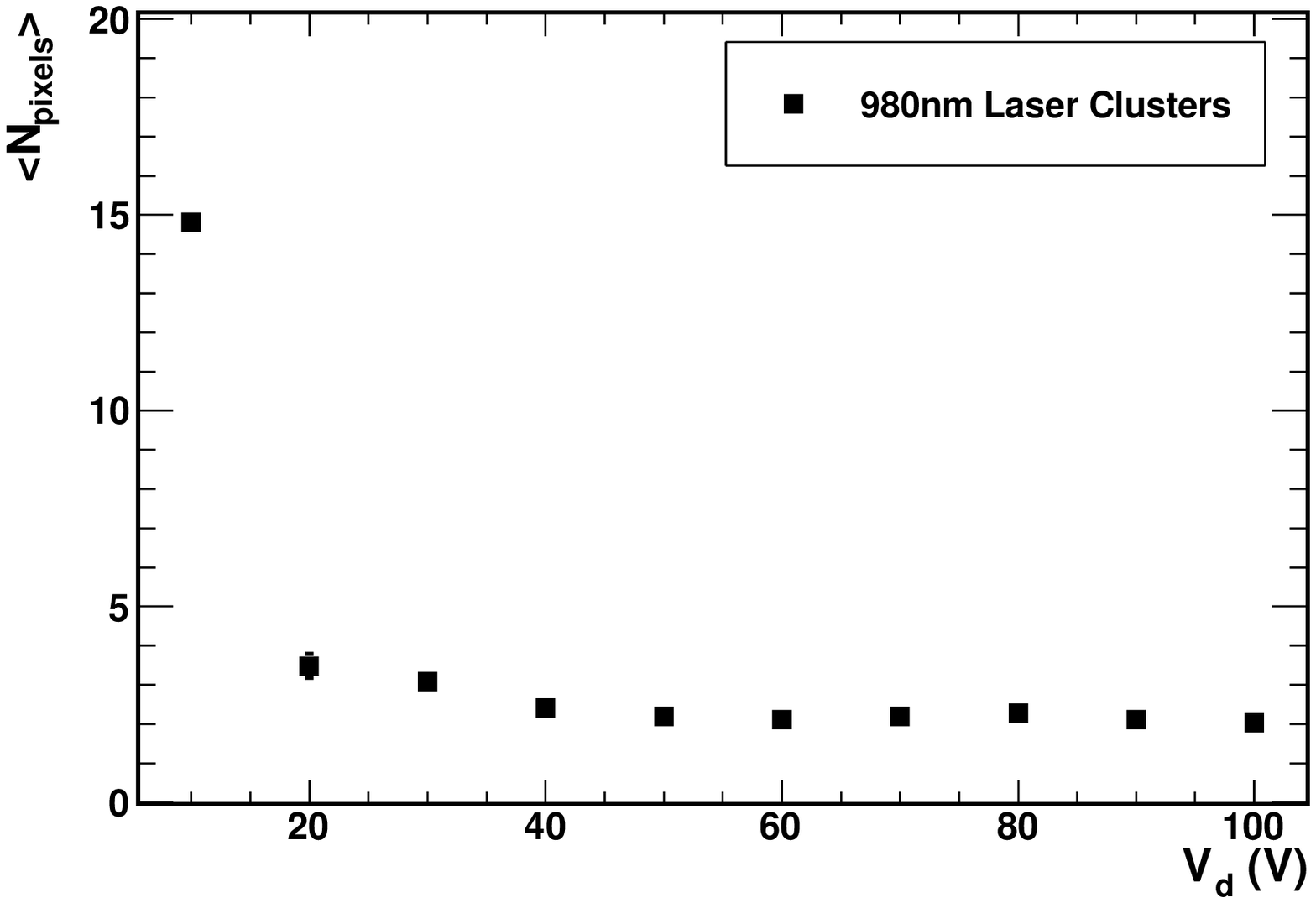}
\includegraphics[width=6.5cm,clip=]{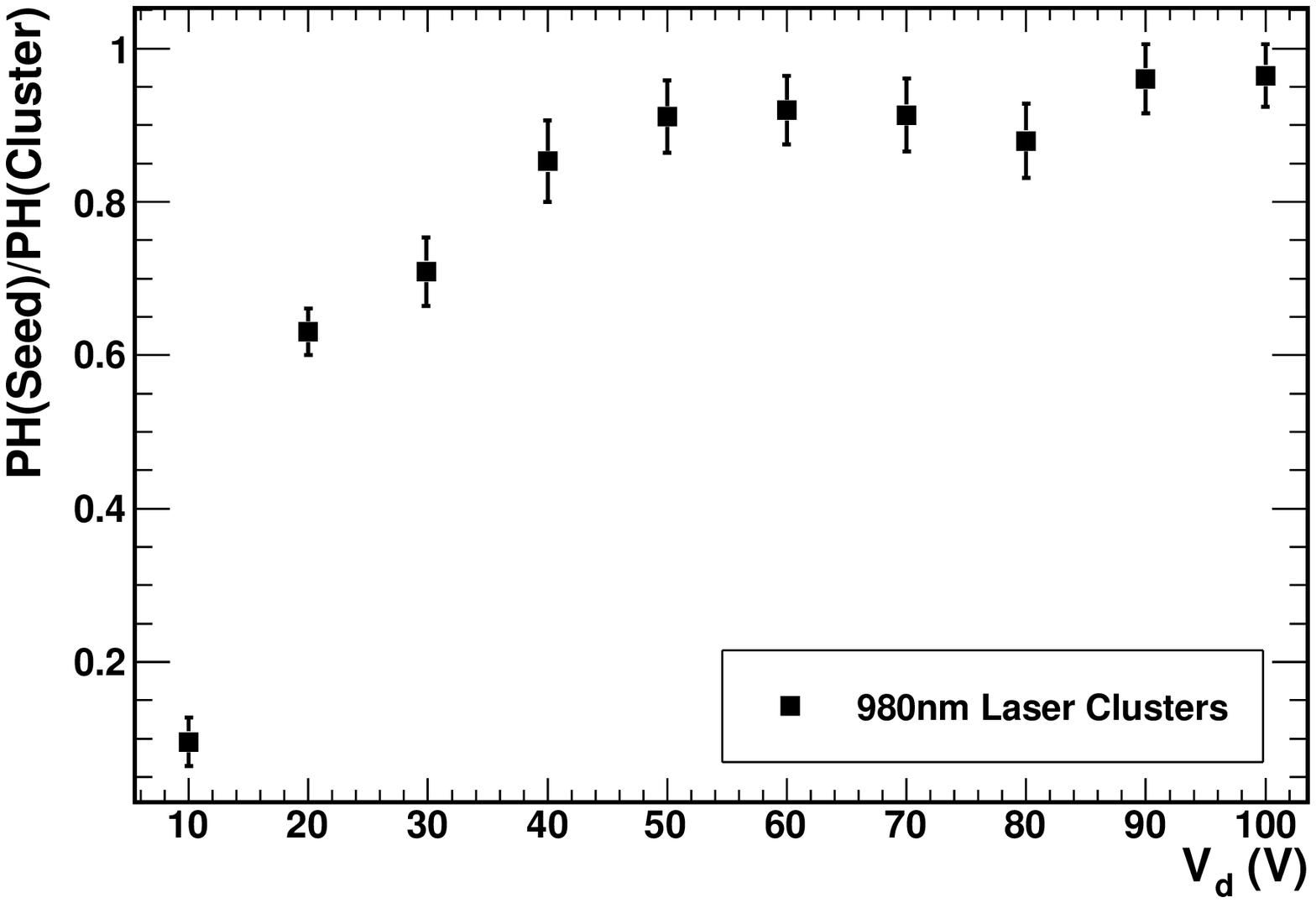}
\end{tabular}
\end{center}
\caption{Average number of pixels in signal clusters and fraction of the total signal in the cluster carried 
by the seed pixel for 980~nm laser pulses in back-illumination configuration as a function of the depletion 
voltage, $V_d$.}
\label{fig:980}
\end{figure}

First the sensor is tested in the laboratory using 2~ns-long pulses of a 980~nm laser collimated to a $\simeq$~5~$\mu$m spot 
on the pixel back-plane. The number of pixels accepted in a cluster, $N_{{\mathrm{pixels}}}$, and the ratio of the signal 
on the seed pixel to the total signal in the cluster measure the spread of the signal charge and the inter-pixel coupling.  
Figure~\ref{fig:980} shows the pixel multiplicity in the clusters and the ratio of the signal on the seed pixel to the 
total signal in the cluster as a function of $V_d$. 
The thinned sensor becomes fully depleted at $\simeq$40~V. We observe that both the pixel 
multiplicity in the clusters, $N_{{\mathrm{pixels}}}$, and the ratio of the seed to cluster charge, $\mathrm{PHR}$, approach a 
plateau with average values of $N_{{\mathrm{pixels}}}$ = 2.16$\pm$0.03 and $\mathrm{PHR}$ = 0.93$\pm$0.01, above $V_d \simeq$40~V.
The full depletion value is confirmed by $C-V$ curves~\cite{soi4}.

The quantum efficiency for X-rays is studied on data collected at the beam-line 5.3.1 of the Advanced Light 
Source (ALS) at LBNL for various values of the photon energy and sensor depletion voltage $V_d$.
The thin, post-processed SOI sensor has been exposed to fluorescence radiation excited in various metal foils, 
with energy in the range 2.1$<E<$8.6~keV. The synchrotron radiation beam from a bending magnet is selected in energy
through a monochromator to obtain a 12~keV monochromatic beam. This beam is sent on a target metal foil at a 45$^{\circ}$ 
incidence angle. A computer-controlled movable stage holds multiple foils, making possible to select different 
target materials remotely. The target and the detector are placed into a vacuum enclosure in order to minimise the 
absorption of soft X-rays in air at the lower energies (below 4~keV) and mounted in back-illumination 
configuration. The angle between the beam and the 
SOI sensor is 90$^{\circ}$. The energy spectrum and the fluorescence radiation intensity are monitored using a 
spectrometer consisting of a Si drift detector~\cite{amptek} with a FWHM resolution of 130~eV at 5.9~keV. This is 
installed at a 30$^{\circ}$ angle w.r.t.\ the incoming beam. The elements used as targets and the energies of 
their dominant fluorescence emission are given in Table~\ref{tab:xrays}. 

\begin{figure}[h!]
\begin{center}
\includegraphics[width=8.5cm,clip=]{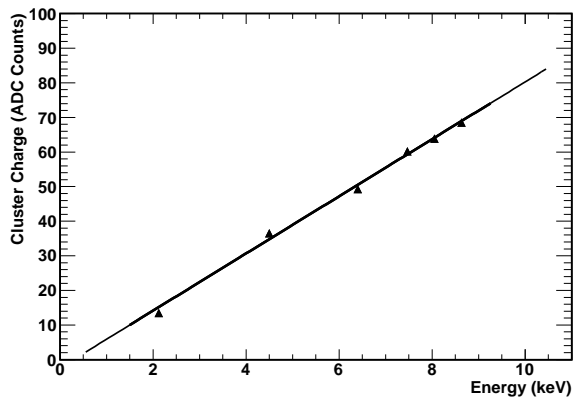}
\end{center}
\caption{Cluster pulse height as a function of the energy of fluorescence X-rays obtained at $V_d$ = 70~V. 
The fitted slope corresponds to a charge-to-voltage conversion of (33.3$\pm$0.4)~$e^-$/ADC count and the 
intercept at zero is -1.7$\pm$0.7~ADC counts.}
\label{fig:cal}
\end{figure} 
\begin{figure}[h!]
\begin{center}
\begin{tabular}{cc}
\includegraphics[width=6.5cm,clip=]{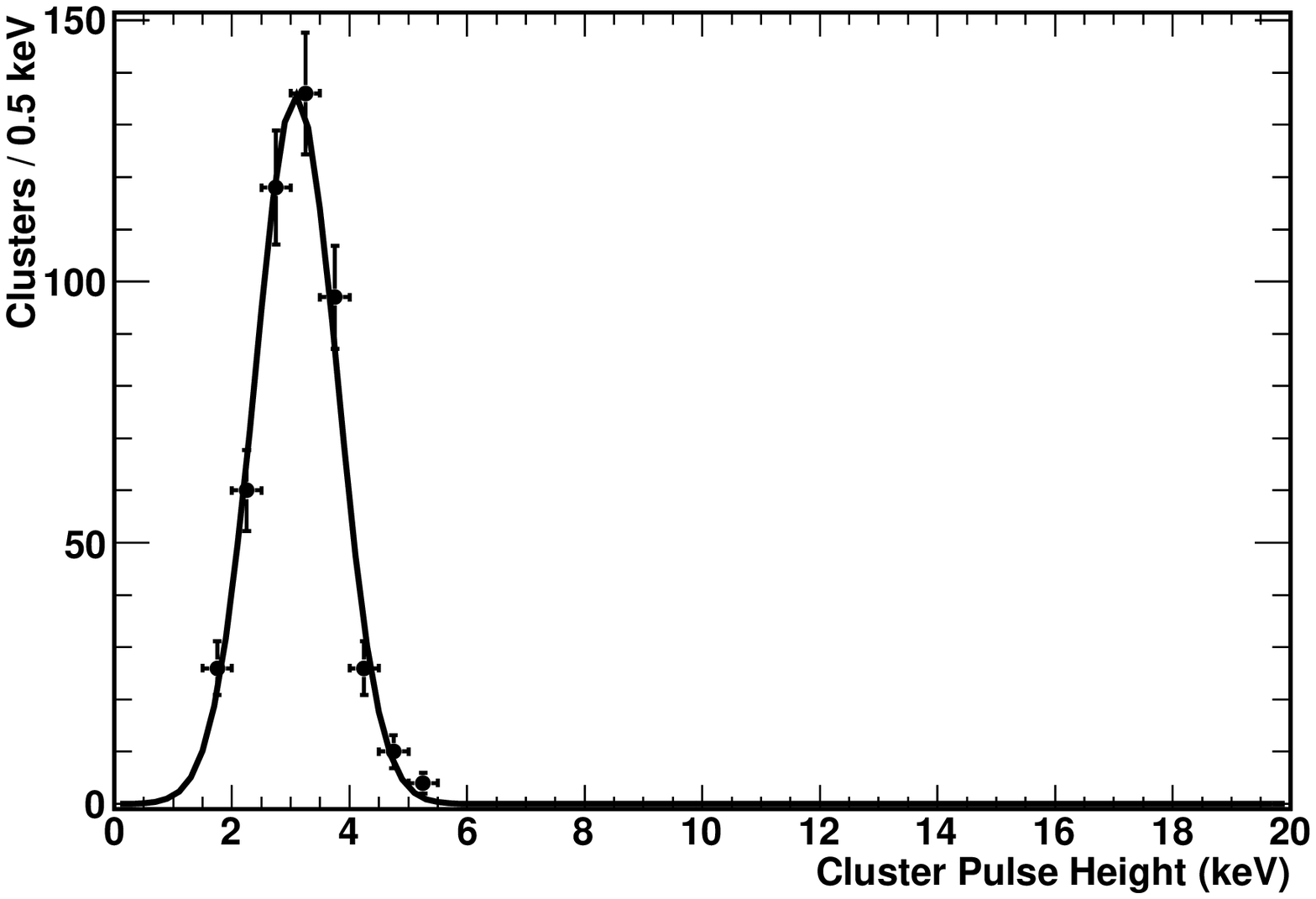} &
\includegraphics[width=6.5cm,clip=]{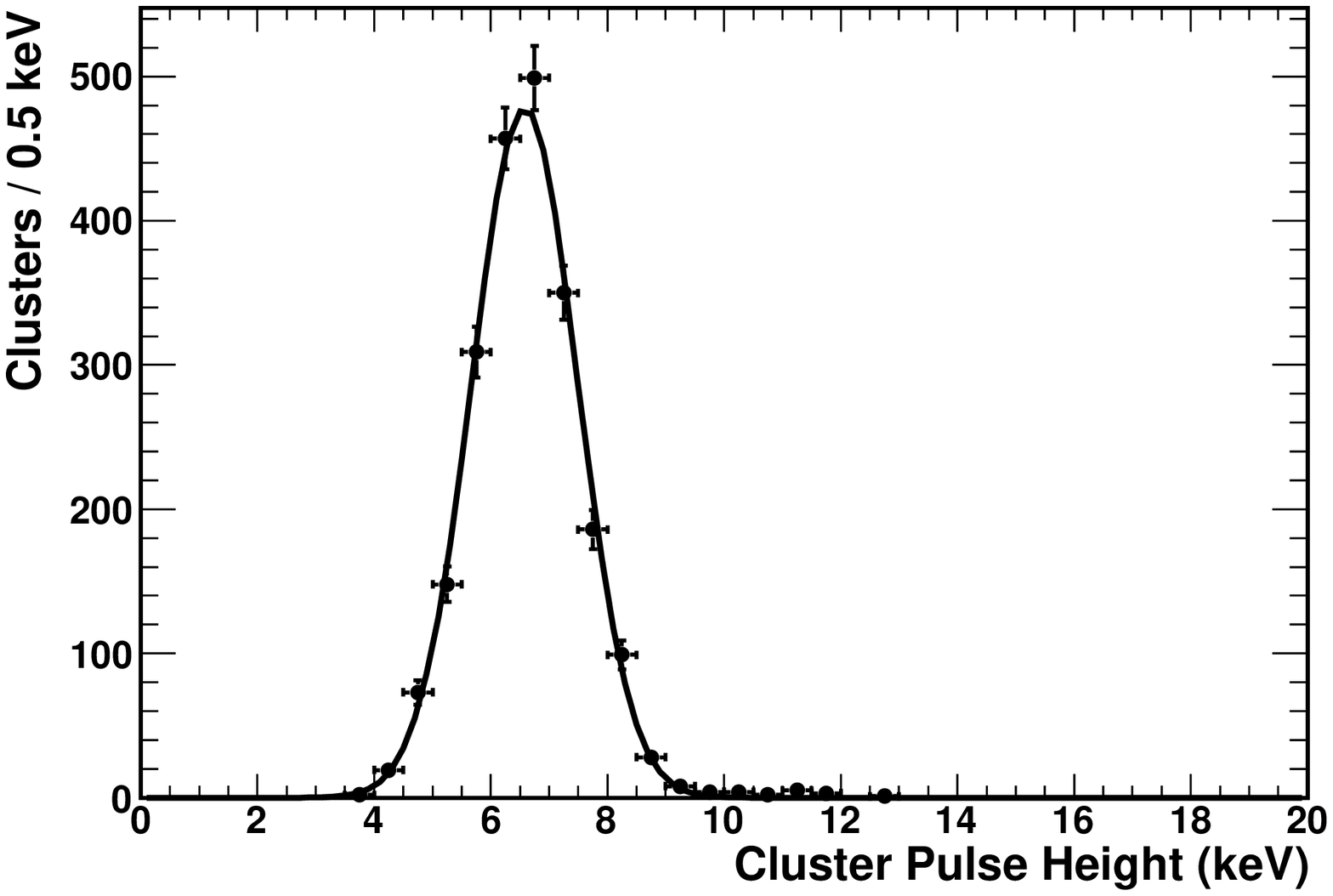}
\end{tabular}
\end{center}
\caption{Cluster pulse height spectrum for Ag (left) and Fe (right) fluorescence obtained at $V_d$ = 70~V. 
The fitted Gaussian width is (0.70$\pm$0.03) and (0.99$\pm$0.02)~keV, respectively.}
\label{fig:spectra}
\end{figure} 
The ADC counts-to-energy conversion is determined from the position of the centroid of the full energy peak 
for X-rays of these energies. We estimate the charge-to-voltage conversion from the slope of the linear fit 
and obtain a calibration of (31.3$\pm$0.4)~$e^-$ ADC count$^{-1}$ at 50~V, consistent with the value obtained 
for a thick, unprocessed sensor of (29.9$\pm$1.2)~$e^-$ ADC count$^{-1}$~\cite{soibeam}, and (33.3$\pm$0.4)~$e^-$ 
ADC count$^{-1}$ at 70~V. 
The calibration at $V_d$ = 70~V is shown in Figure~\ref{fig:cal}. Due to charge diffusion and inter-pixel 
coupling the signal spreads on several pixels. In the energy range of interest for this study, the signal 
cluster size, $N_{{\mathrm{pixels}}}$, varies from 2.02 at 2.98~keV to 3.15 at 8.05~keV as the increasing 
energy deposit increases the likelihood for neighbouring pixels, sharing a constant fraction of the total 
deposited charge, to be above the threshold of 2.5~$\sigma_{\mathrm{noise}}$ used in the cluster 
reconstruction. In order to improve the energy resolution for signal clusters with different 
pixel multiplicity, the energy calibration is performed separately for each value of multiplicity 
1$\le N_{{\mathrm{pixels}}} \le$5. The resulting spectra for Ag and Fe fluorescence are shown in 
Figure~\ref{fig:spectra}. Over the X-ray energy considered, we observe a Gaussian width of the full energy 
peak evolving from 700~eV at 3~keV to 1.14~keV at 8~keV, which is consistent with that expected from the 
measured pixel noise and multiplicity.

\begin{table}
\caption{Summary of the elements used as target and the energy of their dominant fluorescence emission.}
\begin{center}
\begin{tabular}{|c|l|}
\hline
Element & E (keV) \\
\hline
Au      & 2.12 \\
Ag      & 2.98 \\
Ti      & 4.50 \\
Fe      & 6.40 \\
Ni      & 7.47 \\
Cu      & 8.05 \\
Zn      & 8.60 \\
\hline
\end{tabular}
\end{center}
\label{tab:xrays}
\end{table}
The detector quantum efficiency is measured by comparing the number of clusters detected within the full energy 
peak with the rate recorded on the spectrometer. Corrections are applied for the different solid angle, X-ray 
absorption and live exposure times of the SOI sensor and the spectrometer. The effect of the thickness of the 
active volume and window of the spectrometer is corrected using the manufacturer specifications. The efficiency 
of the SOI sensor depends on the product of X-ray absorption in the sensitive thickness and the transmission 
through the back-plane entrance window. We measure the total thickness of the thinned detector, (73$\pm$2)~$\mu$m, 
and estimate the total thickness of the top passivation, the CMOS and the buried oxide layers, to be 9~$\mu$m, from 
manufacturer data. Then, we compute the depletion thickness of the handle wafer assuming the nominal resistivity 
value of 700~$\Omega$$\cdot$cm, which is consistent with the value of (753 $\pm$ 73)~$\Omega$$\cdot$cm determined 
from the charge collected for minimum ionising particles as a function of the depletion voltage~\cite{soibeam}. 
The sensor is fully depleted for the voltages used. The thickness of the entrance window, $d$, is left as a free 
parameter and extracted from the data by a fit. 
\begin{figure}[h!]
\begin{center}
\includegraphics[width=10.5cm,clip=]{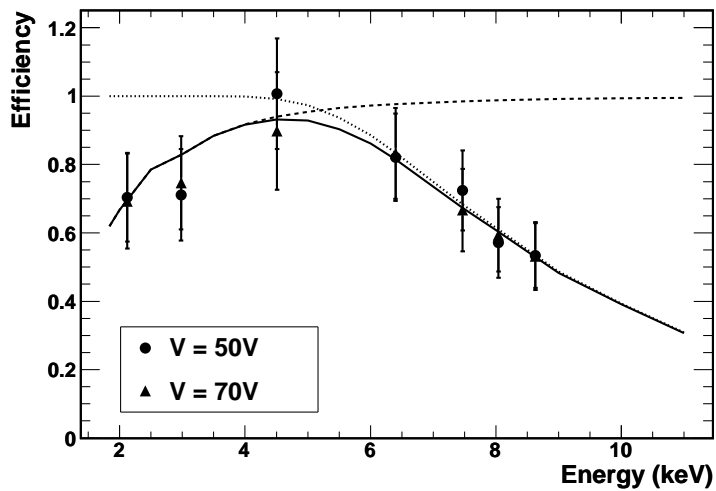}
\end{center}
\caption{Estimated X-ray quantum efficiency as a function of energy for two different values of $V_d$. 
The error bars include the systematic effect due to the system geometry and the normalising of the SOI sensor
counting rates to those measured on the spectrometer and are correlated. The dashed line shows the expected 
efficiency from the absorption in Si, the dotted line that from the transmission through the back-plane entrance window
and the solid line the total efficiency. These curves correspond to the best fit value for the entrance window thickness.}
\label{fig:eff}
\end{figure} 
The back-illuminated thin SOI sensor response to X-rays is studied as a function of the photon energy and 
depletion voltage by comparing the rate of reconstructed clusters, compatible with the full energy peak,  
to the corrected rate recorded on the Si drift spectrometer.

Results are summarised in Figure~\ref{fig:eff}, where the efficiency of the back-illuminated SOI pixel sensor, 
at values of $V_d$ corresponding to full depletion and over depletion, is compared to the expected Si 
absorption efficiency and entrance window transmission. The error bars include the statistical uncertainties 
as well as the systematics coming from the system geometry. The systematic uncertainties are correlated for the 
various energies and voltages. We extract the value of the back-plane window thickness in units of equivalent Si 
thickness by a 1-parameter $\chi^2$ fit to the data and obtain $d$=(0.6$\pm$0.2)~$\mu$m, where the quoted error accounts
for the statistical and systematic uncertainties on the data points. This value is in fair agreement with that of 
the P implant depth of 0.4~$\mu$m obtained from the SRA analysis of the post-processed chip 
(see Figure~\ref{fig:sra}). These results show that the post-processed, fully depleted SOI sensor operates 
with efficiency in excess of 0.60 for X-rays in the range 2$<E<$8~keV. For specific applications, the 
efficiency can be increased in the higher energy end of the spectrum by adopting a higher resistivity handle 
wafer. Recently, pixels have been manufactured on Float Zone (FZ) SOI wafers with resistivity of several 
1000~$\Omega$$\cdot$cm, which can achieve full depletion over a thickness up to 500~$\mu$m.

\section{Potential Applications}

The availability of a pixellated sensor with large quantum efficiency for soft X-rays and low-energy electrons, 
small pixels, fast readout and moderate energy resolution opens up a broad field of potential imaging applications. 
A first domain of possible applications is in precision beam monitoring in linacs and storage rings with X-rays 
and soft electrons. X-ray beam imaging for profile monitoring in low emittance electron storage rings such as 
synchrotron light sources and damping rings has been demonstrated~\cite{fresnel,takano} and it requires a fast 
detector, sensitive to X-rays of a few keV, to minimise diffraction effects, and with small point spread function, 
to minimise resolution effects. Real-time quality control in hadrontherapy requires non-destructive position measurements 
of beams, which can be performed using secondary electron emission~\cite{slim}. In this design the electron beam is 
accelerated to 20~keV and focused onto an imaging detector, for which a post-processed SOI sensor can be an interesting 
solution. Another field where a fully depleted, back-illuminated SOI sensor can be applied to image low-energy electrons 
is in transmission electron microscopy (TEM). Since the displacement damage threshold is proportional to $\sqrt{E}$, it 
is advantageous to use a low energy beam for the study of single atomic layers of carbon in graphene or carbon nano-tubes 
and in biology. In both applications, since the range of 20-40~keV electrons in Si is below 10~$\mu$m, the contribution 
of multiple scattering in the sensor to the point spread function (PSF) is expected to be small. A study of PSF in TEM 
with a front-illuminated CMOS pixel sensor showed that the PSF tends to decrease for electron energies below 120~keV, 
since the decrease of the electron range limits the distance over which charge can be deposited~\cite{emicro}. 
The use of a back-illuminated SOI sensor for low energy TEM has a two-fold advantage. First, the amount of passive material 
traversed by the electron is much reduced compared to front illumination, $\simeq$0.5~$\mu$m compared to $\simeq$4~$\mu$m. 
Second, the radiation damage by the electron beam is greatly reduced, since electrons are stopped in the handle wafer without 
reaching the oxide and the CMOS layers. A detailed simulation study based on {\tt Geant-4} for the TEM application shows 
that the PSF for 20~keV electrons and back-illumination on a thin SOI sensors is comparable to that obtained with 300~keV 
electrons.
  
\section{Conclusions}
A pixel sensor in SOI technology on high-resistivity substrate thinned to 70~$\mu$m to enable full depletion and implanted 
on the back-plane with a thin Phosphor contact, has been characterised with soft X-rays. Its response is evaluated in 
back-illumination using fluorescence radiation in the energy range of 2.12 up to 8.6~keV at the LBNL Advanced Light Source. 
After thinning and back-plane implant the pixel noise and conversion gain are measured and found to be consistent to 
those of sensors before post-processing. Over the X-ray energy range considered, the sensor showed an efficiency in excess 
of 60\%, consistent with the expectations from photoabsorption in Si and transmission through the entrance window. The 
measured quantum efficiency curve is in agreement to that for a sensor with a (0.6$\pm$0.2)~$\mu$m back-plane passive window, 
in units of equivalent Si thickness.

\section*{Acknowledgements}

This work is supported by the Director, Office of Science, of the 
U.S. Department of Energy under Contract No.DE-AC02-05CH11231 and 
by INFN, Italy. We are grateful to Y.~Arai for his effective 
collaboration and support in the SOIPIX activities.

\end{document}